# Simulation studies and spectroscopic measurements of a position sensitive detector based on pixelated CdTe crystals

K. Karafasoulis, K. Zachariadou, S. Seferlis, I. Kaissas, C. Lambropoulos, D. Loukas, C. Potiriadis

*Abstract*–Simulation studies and spectroscopic measurements are presented regarding the development of a pixel multilayer CdTe detector under development in the context of the COCAE project. The instrument will be used for the localization and identification of radioactive sources and radioactively contaminated spots. For the localization task the Compton effect is exploited. The detector response under different radiation fields as well as the overall efficiency of the detector has been evaluated. Spectroscopic measurements have been performed to evaluate the energy resolution of the detector. The efficiency of the event reconstruction has been studied in a wide range of initial photon energies by exploiting the detector's angular resolution measure distribution. Furthermore, the ability of the COCAE detector to localize radioactive sources has been investigated.

## I. INTRODUCTION

THE COCAE instrument is a portable spectroscopic system under development aimed to be used for the accurate localization and identification of radioactive sources and radioactively contaminated spots, in a broad energy range up to 2 MeV.

Among the applications of the instrument are the security inspections at the borders, the detection of radioactive sources into scrap metals at recycling factories, the improvement of the existing procedures at nuclear waste management facilities and the fast and accurate data collection during the response after a nuclear emergency situation.

Manuscript received November 12, 2010. This work was supported by the collaborative Project COCAE SEC-218000 of the European Community's Seventh Framework Program.

K. Karafasoulis is with the Greek Atomic Energy Commission, Agia Paraskevi, Attiki, Greece and with the Hellenic Army Academy, Vari, Attiki Greece (e-mail: ckaraf@gmail.com).

K. Zachariadou was with the Institute of Nuclear Physics, National Center for Scientific Research, Agia Paraskevi, Attiki, Greece. She is now with the Greek Atomic Energy Commission, Agia Paraskevi, Attiki, Greece and with the Technological Educational Institute of Piraeus, Petrou Rali & Thivon-Athens, Greece (e-mail:zacharia@inp.demokritos.gr).

S. Seferlis is with the Greek Atomic Energy Commission, Agia Paraskevi, Attiki, Greece (e-mail: stsefer@eeae.gr).

I. Kaissas is with the Greek Atomic Energy Commission, Agia Paraskevi, Attiki, Greece (e-mail: ikaissas@eeae.gr).

C.P. Lambropoulos is with the Technological Educational Institute of Chalkida, Psahna – Evia, 34400 Greece (e-mail: lambrop@teihal.gr).

D. Loukas is with the Institute of Nuclear Physics, National Center for Scientific Research, Agia Paraskevi, Attiki, Greece (e-mail: loukas@inp.demokritos.gr).

C. Potiriadis is with the Greek Atomic Energy Commission, Agia Paraskevi, Attiki, Greece (e-mail: cpot@eeae.gr).

The COCAE instrument, shown schematically in Fig.1, consists of ten parallel planar layers placed 2cm apart from each other, made of pixelated 2mm thick Cadmium Zinc Telluride (Cd(Zn)Te) crystals occupying an area of 4cmx4cm. Each detector's layer consists of a two dimensional array of pixels (100 x 100) of 400 μm width, bump-bonded on a two dimensional array of silicon readout CMOS circuits of 300 μm thickness. Both pixels and readout arrays are on top of an $Al_2O_3$ supporting printed circuit board layer.

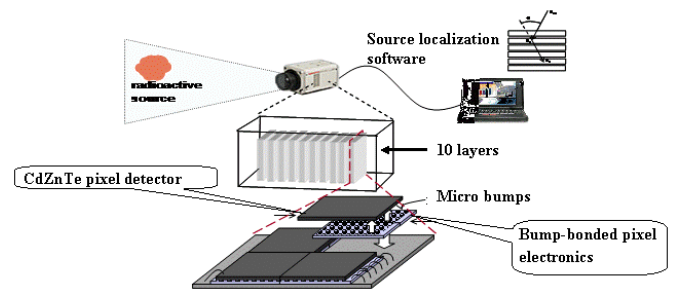

Fig. 1. COCAE instrument conceptual design.

For the source localization task, COCAE exploits the Compton scattering imaging [1], a technique widely used in many fields such as nuclear medicine, astrophysics and recently counterterrorism. Instruments like COCAE that exploit the Compton imaging technique deduce the energy of the incident gamma ray photons as well as their origin within a cone, by measuring the energy depositions and the positions of the Compton scattering interactions recorded in the detector. Successive interactions of an incident photon create an overlapping cone. The intersection of the cones corresponding to different incident photons determines the source location (see Fig.2). In principle, three cones are sufficient to reconstruct the image of a point source. In practice, due to measurement errors and incomplete photon absorption, a large number of reconstructed cones are needed to derive the source location accurately.

A typical design of such an instrument consists of two types of detectors, the scatter detector with relatively low atomic number, where the Compton scattering occurs, and the absorber with relatively high atomic number, in which the scattered photon is ideally totally absorbed.

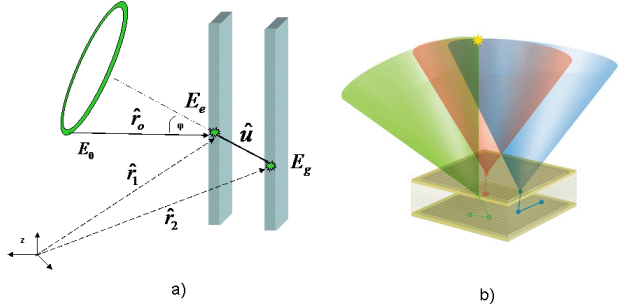

Fig. 2. The principle of Compton reconstruction: a) kinematics of Compton scattering b) illustration of Compton imaging.

The CdTe detectors in the COCAE instrument work both as a scatter, thanks to their arrangement into thin layers, and as an absorber, due to the large atomic numbers of Cd (Z=48) and Te (Z=52), resulting into a high photo-absorption efficiency.

The most important detector parameters are the detection efficiency and the energy resolution, since they mainly influence the event statistics and the uncertainty of the Compton scattering angle determination.

As far as the detection efficiency is concerned, CdTe semiconductor crystals have in principle higher efficiency due to their higher atomic number, compared to Germanium (Ge) and Sodium Iodide (NaI) detectors. In order to achieve even better efficiency, a thick CdTe detector of several mm would be needed, but an increase of the crystal thickness would deteriorate the detector energy resolution due to the effect of incomplete charge collection of CdTe semiconductors. In order to overcome this restriction COCAE instrument is designed as a system of ten 2mm stacked detectors, instead of one thick mono-crystal.

The energy resolution of CdTe detectors is much better compared to NaI, even for the medium quality CdTe crystals. On the other hand the Doppler broadening effect is higher for CdTe detectors compared to NaI and Ge detectors. The challenge for the COCAE instrument is to achieve an energy resolution comparable to the one of high purity Germanium detectors, without the need of cryogenics (CdTe semiconductors can be operated at room temperature due to their high energy bandwidth).

Among the main advantages of the proposed system is its ability to estimate the spatial radioactivity distribution, its enhanced efficiency and its high energy resolution in combination with its independency of any cryogenic system.

Extensive Monte Carlo studies have been performed in the current study in order to:
(a) determine the energy response of the COCAE detector when irradiated with mono-energetic photon fields.
(b) estimate the efficiency of the COCAE detector as a function of the incident photon energy
(c) create a dataset of "hits" (i.e. energy depositions in the detector's pixels and their spatial coordinates) for different incident photon energies, to be used as an input to the Compton reconstruction procedure.

The created dataset will be used for the event reconstruction and specifically to estimate the incident photon direction.

The Compton event reconstruction is performed in two steps:
1. The hits originating from the same incident photon are ordered by means of statistical and physical considerations.
2. The cone that constrains the photon direction is determined using the Compton scattering kinematics.

The first step has been extensively studied in [2].

The main task of the present work is the determination of the direction of the incident photon as well as the source-to-detector distance estimation.

## II. MONTE CARLO SIMULATION

The COCAE detector is modeled with the aid of an open-source object-oriented software library (MEGAlib [3]). For the Monte Carlo simulation, MEGAlib provides interface to the Geant4 [4] simulation toolkit. In Geant4, the Low Energy Compton Scattering (G4LECS [5]) package is used to accurately model the Compton scattering by including Compton cross sections modified for bound electron momentum, as well as corrections on the resulting changes in the scattered particle energies (known as Doppler broadening). The Monte Carlo simulation procedure consists of the following steps:

(a) the description of the exact detector geometry
(b) the generation of photons emitted from an isotropic point source placed at different positions, in an energy range from 60KeV up to 2000KeV. A large number of photons have been generated (~$10^9$) and the energy deposition of each photon interaction ("hit") has been recorded.
(c) the interaction of the generated photons with the detector's materials taking into account all relevant physical processes (Compton scattering, photoelectric effect, pair production, electron/positron transportation into matter, ionization).

### A. Detector Energy Response

In order to adapt the ideal simulated data to realistic energy measurements, the simulated energy depositions are blurred according to Gaussian distributions whose FWHM are determined according to real data. The measurements have been performed using a mono-crystal CdTe Schottky diode, under development within the COCAE project. The diode was biased to 600V negative voltage and illuminated by various radioactive sources. Fig. 3 shows the diode's measured energy resolution (FWHM) to be 3.4% at 60 keV ($^{241}$Am), 2.4% at 122 keV ($^{57}$Co), 2% at 356 keV ($^{133}$Ba) and 1% at 662 keV ($^{137}$Cs).

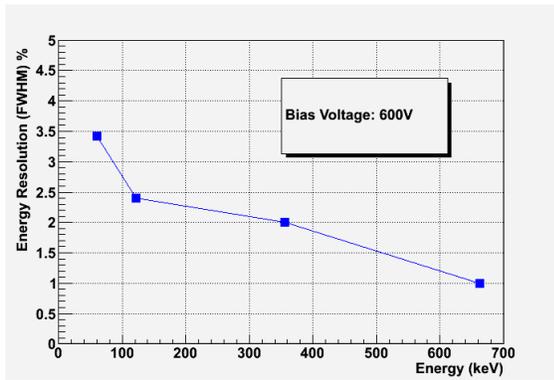

Fig. 3. Photo-peak centroid distribution for the PID350#3 detector illuminated with a $^{109}$Cd radioactive source.

The simulated photon's energy resolution is assumed to be 1% at energies above 662 keV.

Shown in Fig. 4 is the spectrum of the simulated energy deposition on the detector's active volume, of different mono-energetic incident gamma ray energies varying from 200 keV up to 1500 keV.

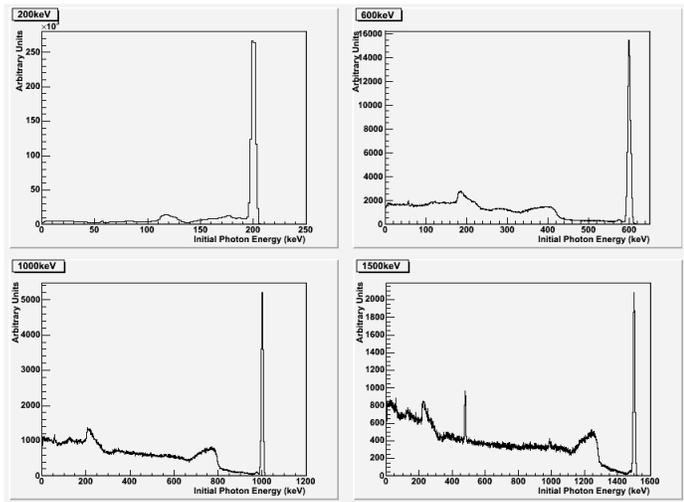

Fig. 4. Typical energy deposition spectra for different gamma ray energies illustrating the photo-peak and the Compton plateau.

Two types of events contribute to the full energy peak of the energy spectrum (Fig 4): (a) events for which the total energy of the photon is transferred to the detector's electrons at once due to photoelectric effect (these are events having one interaction in the detector) and (b) events for which the total energy of the photon is transferred to the detector's electrons after a sequence of Compton scatterings followed by a photoelectric interaction.

The Compton reconstruction is based on the fully absorbed Compton events encountering in the full energy peak.

Fig. 5 summarizes the simulation results for the contribution of multiple cluster (i.e. a collection of neighbor hits) events to the photo-peak. It can be noticed that the number of multiple cluster events contributing to the photo-peak increases as the incident photon energy increases, thus the number of events suitable for reconstruction becomes higher.

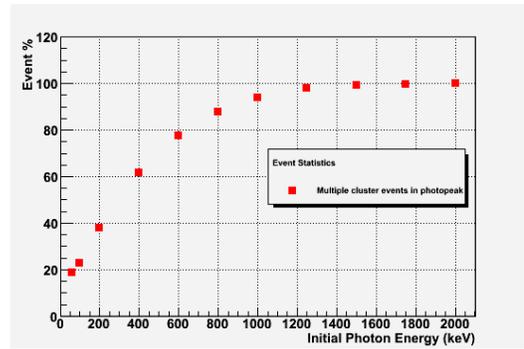

Fig. 5. Percentage of fully absorbed multiple cluster events as a function of the incident gamma ray energy.

### B. Efficiency Studies

The efficiency of the detector has been studied as a function of the incident photon energy.

The absolute efficiency of the COCAE detector calculated for a point source placed at 71 cm from the first detecting layer is plotted in Fig. 6.

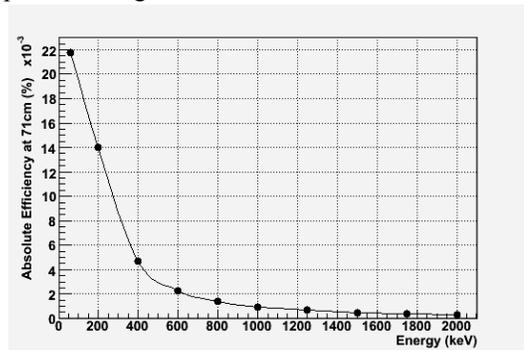

Fig. 6. Detector's absolute efficiency for a point source placed at 71 cm from the first detecting layer and on it's axis of symmetry.

By using various activation patterns of the detector layers, the detector efficiency can be adjusted and the minimum detection limits can be optimized (see Fig. 7). For example, if the low energy part of the spectrum is of interest the first one or two detecting layers can be activated, otherwise all layers remain active.

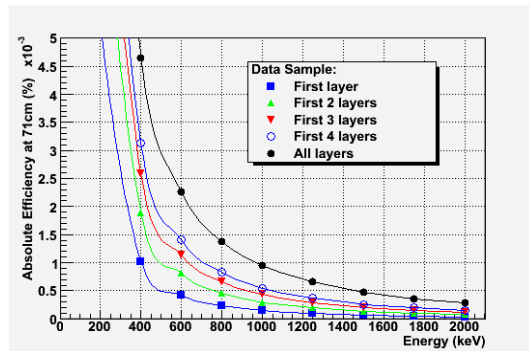

Fig. 7. Detector's absolute efficiency for a point source placed at 71 cm from the first detecting layer and on it's axis of symmetry, when a different number of detector's layers has been activated. The case of all detecting layers being activated is also shown. .

## III. EVENT RECONSTRUCTION

During the event reconstruction process, the spatial and energy information of each individual hit are reconstructed to form events. The reconstructed data are represented by event types (e.g. Compton scattering or photo-effect event) and their associated information (e.g. energy and direction of the Compton scattered gamma ray and of the recoil electron).

For the case of Compton events the reconstruction is split into two steps:

(a) Clustering (blobbing adjacent hits into one larger hit, called from now on an energy cluster or simply a cluster): The energy of the eight closest to a hit neighbor pixels are combined to form a cluster. The energy of the cluster is the total energy of its pixels and its position is calculated via an energy-based center of gravity [6].

(b) Compton sequence reconstruction (identification of the sequence of Compton interactions): the clusters must be sorted in the order in which the interactions of the original particles occurred inside the detector. Since the distance between the layers is very small in a portable instrument such as COCAE, it is impossible to have a timing tag for each hit, thus the interaction sequence cannot be determined directly. The objective of the sequence reconstruction algorithms is to select the most probable cluster sequence among all possible ones and to eliminate the combinatorial, by exploiting the Compton kinematics as well as statistical test criteria.

(c) The COCAE detector's ability to accurately localize radioactive sources depends on the efficient reconstruction of the incident direction of the photons emitted by the source that interact through Compton scattering with the detector. Thus, the determination of the correct sequence of photon interactions in the detector is crucial. The importance of the determination of the correct sequence can be easily visualized in the case of a fully absorbed photon interacting in the active parts of the detector via a Compton scattering (dual cluster event), is illustrated in Fig. 8.

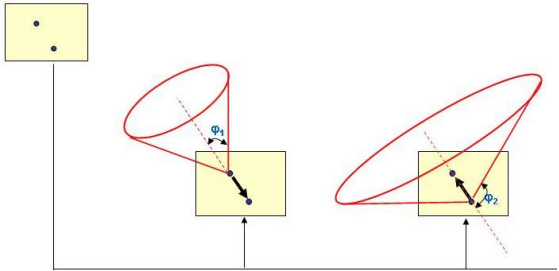

Fig. 8. Compton sequence reconstruction of an event having two interactions in the active volume of the detector. Without time information there are two possible hit orderings resulting to different opening angles of the Compton cone.

### A. Compton Sequence Reconstruction

Different techniques for the Compton sequence reconstruction have been extensively studied in [2] in order to select the best performing one.

Specifically, two different algorithms have been exploited: the Dual Cluster Sequence reconstruction (DCS) applied on events having only two interactions on the detector's active volume and the Multiple Cluster Sequence reconstruction (MCS), applied on events with more than two interactions.

Among the DCS algorithms, two of them have been proved to have the best performance: the DCS-B (that selects the combination with the highest product of the Klein-Nishina differential cross section multiplied by the probability for absorption via photoelectric effect) and the DCS-C (that sorts the clusters assuming that the cluster having the highest energy deposition is the first Compton scattering). Their performance has been evaluated to vary from ~50% up to ~90%. On the other hand the MCS algorithm assigns a quality factor (QF), based on a generalized $\chi^2$ approach [6] to each combination and selects the one with the minimum QF value. The MCS algorithm correctly reconstructs the Compton sequences of more than two cluster events with an efficiency reaching ~55% at energies above 600keV. The MCS algorithm's performance can be improved up to ~70% by applying additional constraints on the minimum value of the quality factor and on the lever arm parameter (the distance between the first and the second interaction), on the cost of lower event statistics.

The efficiency of the best performing Compton sequence reconstruction algorithms is presented in Fig. 9.

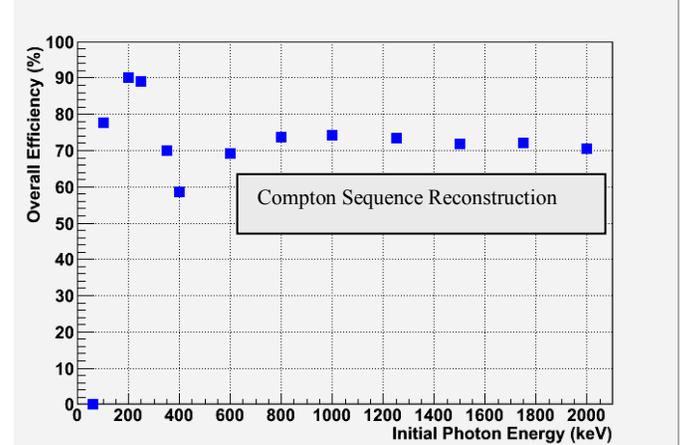

Fig. 9. Compton sequence reconstruction efficiency as a function of the initial photon energy.

### B. Initial photon direction determination

In order to estimate the ability of the detector to evaluate the initial photon direction the Angular Resolution Measure (ARM) is used. The ARM is defined as the angle difference ($\Delta\phi = \phi^{kin} - \phi^{geo}$) between the photon scatter angle $\phi^{kin}$ calculated by the measured energy depositions of the recoil electron and the scattered photon according to the Compton formula and the photon scatter angle $\phi^{geo}$ calculated by the positions ($\vec{r}_1, \vec{r}_2$) of the interactions and the direction $\hat{r}_0$ of the incident gamma ray (see Fig. 10)

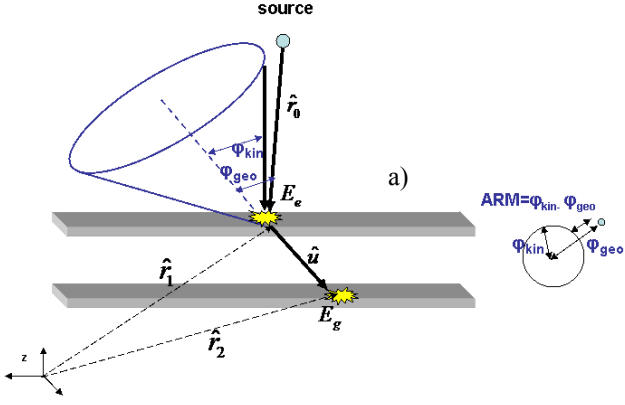

Fig. 10. Definition of the angular resolution measure (ARM).

The ARM distribution is a well-shaped peak (the convolution of a Gaussian and a Lorentz distribution) for low energy photons. Shown in Fig. 16-a is the ARM distribution for the case of 400 keV incident gamma rays, assuming that the detector energy resolution is 5%, taking into account only events having three energy depositions (hits) in the detector's active volume. For higher energies the ARM distribution is bimodal, where the second peak is due to wrongly reconstructed events that are accumulated at large angles (ARM>45$^o$) (Fig. 11-b).

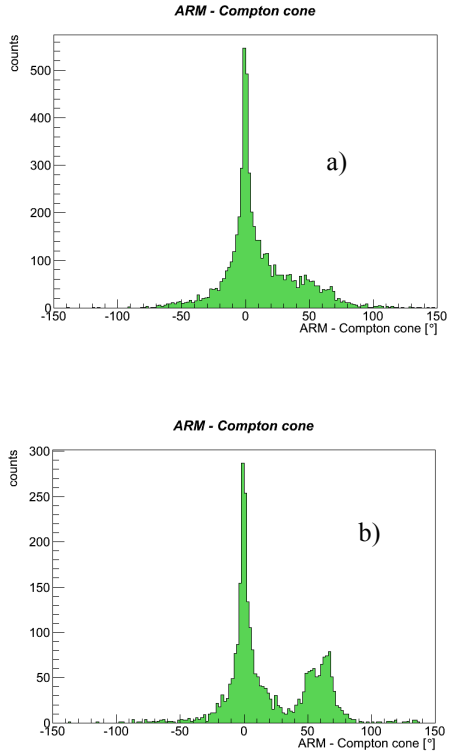

Fig. 11. The angular resolution measure (ARM) distribution for events having three hits for a) 400keV and b) 1000keV incident photon energy

The ARM distribution provides a measure of the performance of the event reconstruction algorithm.

The efficiency of evaluating the direction of the incident photons at a given energy has been defined as the fraction of reconstructed events that lie inside an acceptance interval around the ARM peak. The latter is defined as the interval that contains the 95.5% (2σ) of all the events, assuming perfect event reconstruction.

The ARM acceptance interval varies from ~25$^o$ at 200keV down to ~13$^o$ at 2000keV. The evaluated efficiency as a function of the incident photon energy is shown in Fig. 12.

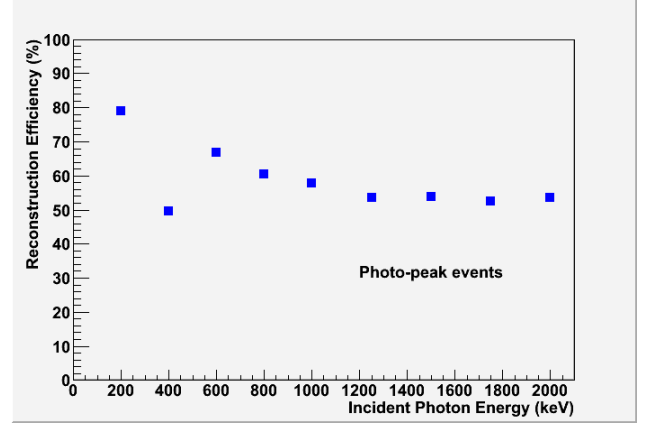

Fig. 12. Efficiency of evaluating the direction of the incident photons as a function of their energy.

## IV. SOURCE LOCALIZATION

The ability of the COCAE detector to localize a radioactive source has been investigated. The source localization task includes the source-to-detector distance as well as the source direction estimation.

The estimation of the source-to-detector distance (i.e. the distance of the source form the first detecting layer) is based on the number of the fully absorbed photons in each detector layer.

The number of photo-peak counts ($N_i$) from each COCAE detector layer (i), is evaluated by fitting the following function:

$$N_i \propto \exp\left(-(i-1)\left(\sum_j \mu_j t_j\right)\right) \cdot \left(\frac{z}{z+(i-1)g}\right)^2 \quad (1)$$

where $t_j$ is the thickness of a material of each detecting layer with corresponding total absorption coefficient $\mu_j$, g is the distance between the layers and z the source-to-detector distance determined by the fitting procedure. The sum runs over all materials of the i$^{th}$ layer.

The estimated source-to-detector distance as a function of the incident gamma rays energy, is presented in Fig. 13, for a point source placed at z=71 cm (Fig.13-a) as well as at z=111cm (Fig 13-b), from the first detecting layer. The large error bars in the case of z=111 cm are due to limited statistics.

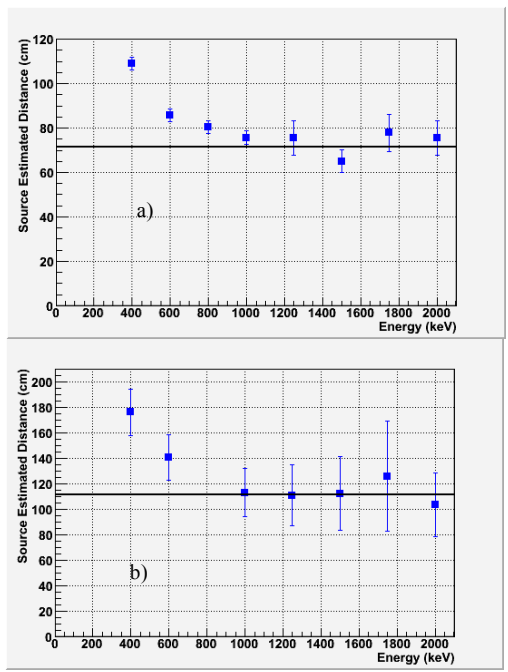

Fig. 13. Estimated distance as a function of the initial photon energy, for a point source placed at a) $z=71$ cm and b) $z=111$ cm form the first detecting layer .

The evaluation of the source direction is performed by using the List Mode Maximum Likelihood Expectation Maximization (LM-ML-EM) imaging algorithm [7].

The reconstructed image of a simulated point source with isotropic emission of 400 keV photons, placed on the symmetry axis of the detector and at 50 cm distance from its centre is shown in Fig. 14.

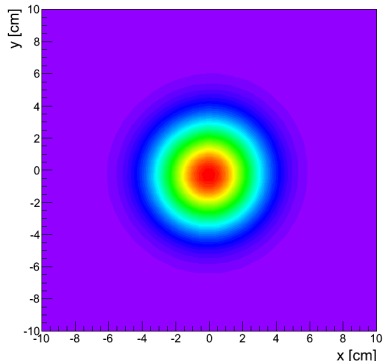

Fig. 14. Reconstructed image of a 400 keV point source located at 50 cm from the COCAE detector, using the LL-ML-EM image reconstruction algorithm.

## V. Conclusion

Simulation studies of a position sensitive portable detector (COCAE) based on pixelated CdTe crystals, are presented.

The exact geometry of the COCAE instrument is modelled with the aid of the MEGAlib library, providing interface to the Geant4 simulation toolkit. Different algorithms for the Compton sequence reconstruction have been reported.

Extensive simulation studies have been performed to understand the detector response under different radiation fields.

The efficiency of the detector has been studied as a function of the incident photon energy.

The Angular Resolution Measure (ARM) distribution has been studied since it provides a measure of the event reconstruction algorithm performance. The latter has been evaluated in a wide range of initial photon energies to vary from around 80% at 200 keV down to ~55% at energies above 800 keV. The ARM FHWM varies from ~25° at 200 keV down to ~13° at 2000 keV.

The ability of the COCAE detector to localize a radioactive source has been evaluated by using algorithms that determine the source-to-detector distance as well as the direction of the source. The source-to-detector distance has been accurately determined within errors for incident photon energies varying from 800 keV up to 2000 keV, for distances up to 111 cm.

The discrepancies observed in the low energies are currently under investigation.

## References

[1] T.Kamae, R. Enomoto, and N. Hanada, "A new method to measure energy, direction, and polarization of gamma rays", *Nuclear Instruments & Methods in Physics Research*, A260 (1987) 254-257
[2] K. Karafasoulis, K. Zachariadou, C. Potiriadis, S. Seferlis, I.Kaissas, D. Loukas, C. Lambropoulos "Evaluation of Compton scattering sequence reconstruction algorithms for a portable position sensitive radioactivity detector based on pixelated Cd(Zn)Te crystals" *arXiv* 1011.2604, http://arxiv.org/abs/1011.2604, submitted for publication.
[3] Zoglauer, R. Andritschke, F. Schopper: "MEGAlib-The Medium Energy Gamma-ray Astronomy Library", New *Astronomy Reviews*, Volume 50, Issues 7-8, Pages 629-632, October 2006
[4] GEANT4-A toolkit for the simulation of the passage of particles through matter, http://geant4.web.cern.ch/geant4/
[5] G4lecs:A Low-Energy Compton Scattering Package, http://public.lanl.gov/mkippen/actsim/g4lecs/
[6] Andreas Christian Zoglauer, "First Light for the Next Generation of Compton and Pair Telescopes", Ph.D. Thesis, Max-Planck-Institute, 2005
[7] Wilderman, S. et al. "List-mode Maximum Likelihood Reconstruction of Compton Scatter Camera Images in Nuclear Medicine", IEEE Trans. Nucl. Sci. 45:957,1988